\documentclass[pra,twocolumn,amsmath,amssymb,showpacs]{revtex4}

\usepackage{graphics}

\usepackage{epsfig}

\usepackage{bbm}

\newcommand{\bra}[1]{\ensuremath{\left\langle{#1}\right\vert}}

\newcommand{\ket}[1]{\ensuremath{\left|{#1}\right\rangle}}
\newcommand{\ad}{\ensuremath{a^\dagger}}
\def\be{\begin{equation}}
\def\ee{\end{equation}}
\def\eea{\end{eqnarray}}
\def\bea{\begin{eqnarray}}

\newcommand{\va}[1]{\ensuremath{(\Delta#1)^2}}

\newcommand{\ex}[1]{\ensuremath{\left\langle{#1}\right\rangle}}

\newcommand{\exs}[1]{\ensuremath{\langle{#1}\rangle}}
\newcommand{\eins}{\ensuremath{\mathbbm 1}}

\begin{document}
\title{Entanglement Witnesses in Spin Models}
\date{\today}

\author{G\'eza T\'oth}

\affiliation{Theoretical Division, Max Planck Institute
for Quantum Optics, Hans-Kopfermann-Stra{\ss}e 1, D-85748 Garching, Germany}

\begin{abstract}
We construct entanglement witnesses using fundamental quantum
operators of spin models which contain two-particle interactions
and have a certain symmetry. By choosing the Hamiltonian as such
an operator, our method can be used for detecting entanglement by
energy measurement. We apply this method to the Heisenberg
model in a cubic lattice with a magnetic field, the XY model
and other familiar spin systems.
Our method provides a temperature bound for
separable states for systems in
thermal equilibrium.
We also study the Bose-Hubbard model and
relate its energy minimum for separable
states to the minimum obtained from
the Gutzwiller ansatz.
\end{abstract}



\maketitle

\section{Introduction}

Entanglement lies at the heart of quantum mechanics
and plays also an important role in the
novel field of Quantum Information Theory (QIT, \cite{BE00}).
While for pure quantum states it is equivalent to correlations, for
mixed states the two notions differ.
In this general case, a quantum state is entangled if its density matrix
cannot be written as a convex sum of product states.
Based on this definition, several sufficient
conditions for entanglement have been developed \cite{BE00}.
In special cases, e.g.
for $2\times 2$ (two-qubit) and $2\times 3$ bipartite systems \cite{AP96} and
for multi-mode Gaussian states \cite{GK01}
even necessary and sufficient conditions are known.

However, in an experimental situation usually only limited information
about the quantum state is available.
Only those approaches for entanglement
detection can be applied which require the
measurement of not too many observables.
One of such approaches is using entanglement witnesses.
They are observables which have a positive expectation value
or one that is zero for all separable states.
Thus a negative expectation value signals the
presence of entanglement.
The theory of entanglement witnesses has recently been rapidly developing
\cite{H96}.
It has been shown how to generate entanglement witnesses that detect
states close to a given one,
even if it is mixed or a bound entangled state \cite{AB01}.
It is also known how to optimize a witness operator
in order to detect the most entangled states \cite{LK00}.

Beside constructing entanglement witnesses, it is also important
to find a way to measure them. For example, they can easily be measured
by decomposing them into a sum of locally measurable terms \cite{BE03}.
In this paper we follow a different route.
We will construct witness operators of the form
\be
W_O:=O-\inf_{\Psi \in S} \big[ \exs{\Psi|O|\Psi}\big],
\ee
where $S$ is the set of separable states,
"$\inf$" denotes infimum, and
$O$ is a fundamental quantum operator of a spin system
which is easy to measure.
In the general case $\inf_{\Psi \in S} \exs{\Psi|O|\Psi}$
is difficult, if not impossible, to
compute \cite{EH04}. Thus we will concentrate
on operators that contain only two-particle interactions
and have certain symmetries.
We derive a general method to find
bounds for the expectation value of such operators for separable states.
This method will be applied to spin lattices. We will also consider
models with a different topology.

If observable $O$ is taken to be the Hamiltonian then
our method can be used for detecting entanglement
by energy measurement \cite{BV04}.
While our approach does not require that the system
is in thermal equilibrium, it can readily be used
to detect entanglement for a range of well-known systems
in this case.
The energy bound for
separable states correspond to a temperature bound.
Below this temperature the
thermal state is necessarily entangled.
Numerical calculations have been carried out for
some familiar spin models.
They show that
for the parameter range in which
substantial entanglement is
present in the thermal ground state,
our method detects the state as entangled.
Thus our work contributes to recent efforts
connecting QIT and the statistical physics
of spin models \cite{SPINMODELS}.

\section{ Energy bound for separable sates}

We consider
a general observable $O$ on a spin lattice defined in terms of
the Pauli spin operators $\vec{\sigma}^{(k)}=
(\sigma_x^{(k)},\sigma_y^{(k)},\sigma_z^{(k)})$ as \be
O:=\mathcal{O}\left[\{\vec{\sigma}^{(k)}\}_{k=1}^N\right],
\label{Odef}\ee where $\mathcal{O}$
is some multi-variable function \cite{QUANTUMCLASICAL}. We will
discuss how to find the minimum expectation value of such
an operator for separable states of the form
\begin{equation}
\rho = \sum_l p_l \rho_l^{(1)} \otimes \rho_l^{(2)} \otimes
...\otimes \rho_l^{(N)}. \label{sep}
\end{equation}

The minimum of $\exs{O}$ for pure product states is obtained
by replacing the Pauli spin matrices
by real
variables $s_{x/y/z}^{(k)}$ in Eq.~(\ref{Odef}) and minimizing it
with the constraint that $\vec{s}^{(k)}$ are unit vectors
\cite{SPIN1}. The minimum
obtained this way is clearly valid also for mixed separable states
since the set of separable states is convex
\be
O_{\text{sep}}:=\inf_{\Psi \in S} \exs{\Psi|O|\Psi}=
\inf_{\{\vec{s}^{(k)}\}}
\mathcal{O}\left[\{\vec{s}^{(k)}\}_{k=1}^N\right]. \label{Ems} \ee

In the most general case many-variable minimization is needed for
obtaining $O_{\text{sep}}$. In some cases, to which many of the most
studied lattice Hamiltonians belong, it is possible to find a
simple recipe for computing the minimum of $\mathcal{O}$.

(i) Let us consider an operator $O$ which is the sum of two-body
interactions. It can be described by a lattice or a graph. The
vertices $V:=\{1,2,...,N\}$ correspond to spins and the edges
between two vertices indicate the presence of interaction.

\begin{figure}
\centerline{\epsfxsize=2.3in\epsffile{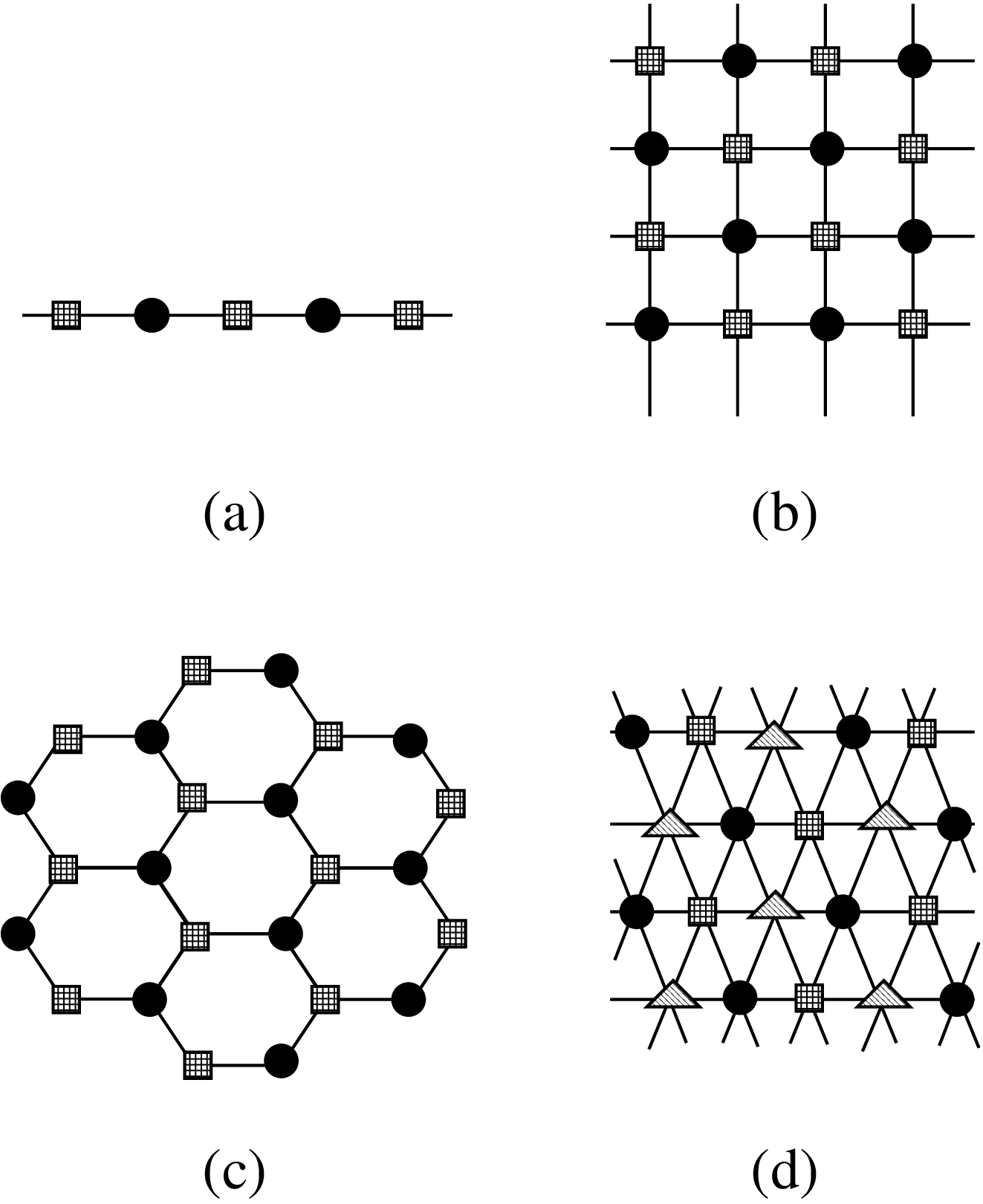}}
\caption{Some of the most often considered lattice models:
(a) Chain, (b) two-dimensional
cubic lattice, (c) hexagonal lattice
and (b) triangular lattice.
Different symbols at the vertices
indicate a possible partitioning into sublattices.}
\label{fig_latt}
\end{figure}

(ii) Let us assume that this lattice can be partitioned into
sublattices in such a way that interacting spins correspond to
different sublattices. Fig.~\ref{fig_latt} shows lattices of
some common one- and two-dimensional spin models.
The different symbols at the vertices
indicate a possible partitioning into sublattices with the above property.
For simplicity, next we will consider
the case with only two disjoint sublattices,
$A$ and $B$, and assume that $\mathcal{O}$ can be written in
the form \be
\mathcal{O}\left[\{\vec{s}^{(k)}\}_{k=1}^N\right]=\sum_{\vec{s}_k^A\in
A,\vec{s}_k^B\in B} f(\vec{s}_k^A,\vec{s}_k^B) \label{Htilde} \ee
where $f$ is some two-spin function, and $\vec{s}_k^{A/B}$
denotes spins of sublattice $A/B$.

If conditions (i) and (ii) are met, then it is enough to find
spins $\vec{s}^A$ and $\vec{s}^B$ corresponding to the minimum of
$f(\vec{s}^A,\vec{s}^B)$. Then setting all the spins in sublattice
A to $\vec{s}^A$ and in sublattice B to $\vec{s}^B$, respectively,
gives a solution which minimizes $\mathcal{O}$.

\section{Examples}

In the following we
will use the Hamiltonian $H$ for constructing entanglement witnesses.
The energy minimum for separable states is the same as the
ground state of the corresponding classical spin model. Our method
detects entanglement if \be \Delta E:=\exs{H}-E_{\text{sep}}<0.
\label{Delta} \ee If $\Delta E<0$ then $|\Delta E|$ characterizes
the state of the system from the point of view of the robustness
of entanglement. It is a lower bound on the energy that the
system must receive to become separable.

We will use the previous results to detect entanglement in
thermal states of spin models. In thermal equilibrium the state of
the system is given $\rho_T=\exp(-H/k_BT)/Tr[\exp(-H/k_BT)]$
where $T$ is the temperature and $k_B$ is the Boltzmann constant.
For simplicity we will set $k_B=1$. Using
Eq.~(\ref{Delta})  a temperature bound, $T_E$, can be found such that
when $T<T_E$ then the system is detected
as entangled.

\subsection{Heisenberg lattice}

Let us consider
an anti-ferromagnetic Heisenberg
Hamiltonian with periodic boundary
conditions on a $d$-dimensional cubic lattice
\be
H_H:=\sum_{\langle k,l\rangle} \sigma_x^{(k)}\sigma_x^{(l)}+
\sigma_y^{(k)}\sigma_y^{(l)}+\sigma_z^{(k)}\sigma_z^{(l)}+B\sigma_z^{(k)}.
\label{HH}
\ee
The strength of the exchange interaction is
set to be $J=1$, $B$ is the magnetic field,
and $\langle k,l \rangle$ denotes spin pairs
connected by an interaction. The expectation value of
Eq.~(\ref{HH}) for separable states is bounded from below \be
\exs{H_{H}}\ge E_{H,\text{sep}}:= \bigg\{
\begin{tabular}{ll}
$-dN[(B/d)^2/8+1]$& if $|B/d|\le 4$, \\
$-dN(|B/d|-1)$& if $|B/d| > 4$, \label{HHb}
\end{tabular}
\ee where $N$ is the total number of spins.
This bound was obtained using two sublattices, minimizing
the expression
$f_H(\vec{s}^A,\vec{s}^B):=\vec{s}^A\vec{s}^B+
B(s_z^{A}+s_z^{B})/(2d)$. Based on this $E_{H,\text{sep}}=dN\inf[
f_H]$ \cite{ODDEVEN}.

Let us now consider a one-dimensional spin-$1/2$ Heisenberg chain
of even number of particles.
If $B=0$ then Eq. (\ref{HHb}) corresponds to
\be
\frac{1}{N}\sum_{\langle k,l\rangle}
\exs{\vec{\sigma}^{(k)}\vec{\sigma}^{(l)}}
\ge -1.
\ee
which is simply a necessary condition for separability in terms of
nearest-neighbor correlations.
In the large $N$ limit, the energy minimum for
entangled states can be obtained as $E_{min}=
-4N(\ln 2-1/4) \approx -1.77N$ \cite{T99}. The
energy gap between the minimum for separable states and the ground
state energy of $H_H$ is thus $\Delta E_{gap} \approx 0.77 N$ which
increases linearly with the number of spins.
As shown in Refs.~\cite{CW01,W02}, when $B=0$
the concurrence of the two-qubit reduced density matrix
in the thermal state is obtained as
$C=\max[-(\exs{H_H}/N+1)/2,0]$. Hence
$C>0$ if $\exs{H_H}<-N$ and $E_{H,\text{sep}}$ coincides with the
energy bound for nonzero concurrence.

\begin{figure}
\centerline{\epsfxsize=1.7in\epsffile{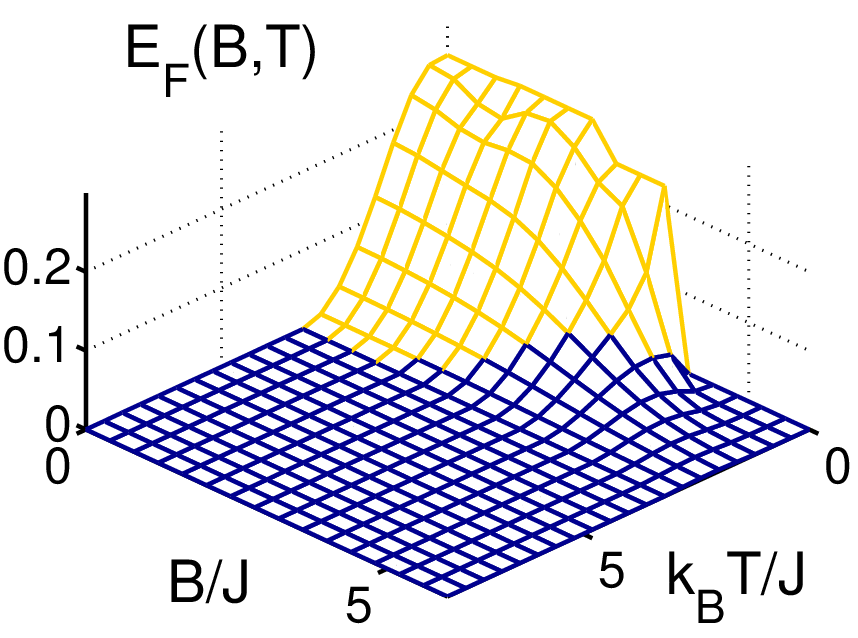}
\epsfxsize=1.8in\epsffile{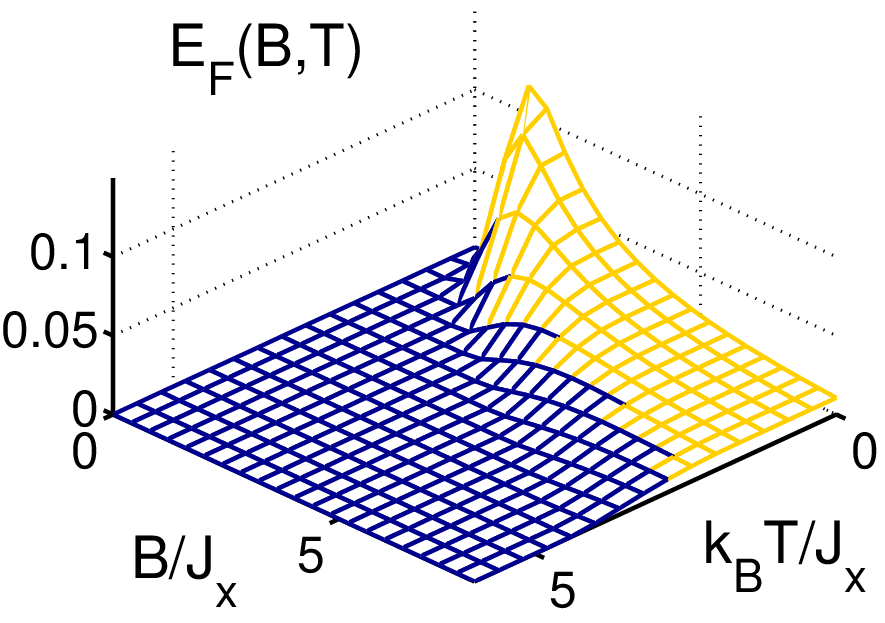}}
(a)\hskip4.0cm (b)
\caption{(a) Heisenberg chain of $8$ spins.
Nearest-neighbor entanglement as a function of magnetic field $B$ and
temperature $T$.
(b) The same for an Ising spin chain.
Here $k_B$ is the Boltzmann constant, $J$ and $J_x$ are coupling constants.
Light color indicates the region where entanglement is detected by our
method.}
\label{fig_Heisenberg}
\end{figure}

Let us now consider the case $B>0$.
Fig.~\ref{fig_Heisenberg}(a)
shows the nearest-neighbor entanglement vs. $B$ and $T$.
The entanglement of formation was computed from the concurrence
 \cite{BV96}. Light
color indicates the region where the thermal ground state is
detected as entangled. There are regions with $C>0$ which are
not detected. However, it is clear that
when the system contains at least a small amount of entanglement
($\sim 0.07$) the state is detected as entangled. Note that
the sharp decrease of the nearest-neighbor entanglement
around $B_{crit}=4$ for $T=0$ is due to
 a {\it quantum phase transition}.

Another important question is how the temperature bound $T_E$
depends on the number of particles.
For the
Heisenberg model of even number of spins
with $B=0$ this temperature decreases slowly with
$N$ and saturates at $T\approx 3.18$.
Ref.~\cite{W02} finds the same bound temperature
for nonzero concurrence for an
infinite system.

\subsection{XY model}

For the XY Hamiltonian on a
$d$-dimensional cubic lattice with
periodic boundary conditions \be H_{XY}:=\sum_{\langle k,l\rangle}
J_x{\sigma_x^{(k)}\sigma_x^{(l)}}+J_y{\sigma_y^{(k)}\sigma_y^{(l)}}
+B \sum_k^{N}\sigma_z^{(k)} \ee the energy of separable states is
bounded from below \be \exs{H_{XY}}\ge E_{XY,\text{sep}}:= \bigg\{
\begin{tabular}{ll}
$-dNM\big(1+b^2/4\big)$& if $b\le 2$, \\
$-dNMb$& if $b> 2$.
\end{tabular}
\ee Here $J_{x/y}$ is the nearest-neighbor
coupling along the $x/y$ direction, $B$ is the magnetic field,
$M:=\max(|J_x|,|J_y|)$ and $b:=|B|/M/d$.
This bound is simply the mean-field ground
state energy if both $J_x$ and $J_y$ are negative. It was obtained using two sublattices and minimizing
$f_{XY}(\vec{s}^A,\vec{s}^B):=J_xs_x^As_x^B +J_ys_y^As_y^B
+B(s_z^A+s_z^B)/(2d)$.

A one-dimensional spin-$1/2$ Ising chain is a special
case of an XY lattice with $J_x=1$ and $J_y=0$.
Fig.~\ref{fig_Heisenberg}(b)
shows the nearest-neighbor entanglement as a function of $B$ and
$T$ for this system.
According to numerics, $T_E$ (computed for $B=1$) decreases
with increasing $N$.
For $N=\infty$ we obtain $T_E\approx 0.41$  \cite{P70}.

\subsection{Heisenberg coupling between all spin-pairs}

From a
theoretical point of view, it is interesting to consider
a system in which the interactions are described
by a complete graph rather than
a lattice \cite{W04}.
For the
following Hamiltonian of $N$ spin-$1/2$ particles \be
H_S:=J_x^2+J_y^2+J_z^2 \ee the expectation value for separable
states is bounded \cite{HT02} \be \exs{H_S} \ge E_{S,\text{sep}}:=2N.
\label{Smin} \ee Here $J_{x/y/z}=\sum_k
\sigma_{x/y/z}^{(k)}$ and for simplicity $N$ is taken to be even.

Now we could not use the method for partitioning the spins into
sublattices. The proof of Eq.~(\ref{Smin}) is based on the theory
of entanglement detection with uncertainty relations \cite{HT02,G03}.
For separable states one obtains \cite{HT02}
\bea
&&\va{J_x}+\va{J_y}+\va{J_z}\nonumber\\
&&\ge \sum_l p_l
\sum_k \bigg[\va{\sigma_x^{(k)}}_l + \va{\sigma_y^{(k)}}_l
+ \va{\sigma_z^{(k)}}_l\bigg] \nonumber\\ &&\ge N \cdot L_S,
\eea
where index $l$ denotes the $l$-th subensemble and
$L_S=\inf_\Psi[ \va{\sigma_x}_\Psi+\va{\sigma_y}_\Psi+\va{\sigma_z}_\Psi]=2$.
Hence Eq. (\ref{Smin}) follows.
The measured energy even gives information
on the entanglement properties of the system.
Based on the previous considerations,
it can be proved that for pure states
$\exs{H_S}/2$ is an upper bound
for the number of unentangled spins, 
$N_{u}$. For mixed states of the form
$\rho=\sum_k p_k \ket{\Phi_k}\bra{\Phi_k}$
we obtain $\exs{H_S}/2 \ge \sum_k p_k N_{u,k}$.
Here $N_{u,k}$ corresponds to the $k$th 
pure subensemble.

Following the approach of Ref. \cite{CW01},
the concurrence
can be computed as a function of the energy.
For even $N$ the concurrence is
$C=\max\{-[\exs{H_S}+N(N-4)]/[2N(N-1)],0\}$.
Since for all quantum states
$\exs{H_S}\ge 0$, the concurrence is zero for
any temperature if $N\ge 4$.
Thus our condition can detect multi-qubit
entanglement even when two-qubit entanglement is not present.

The thermodynamics of $H_S$ can be obtained by knowing the energy
levels and their degeneracies \cite{CE99} \bea E_j&=&2j(2j+2),\nonumber\\
d_j&=&\frac{(2j+1)^2}{N/2+j+1} \binom{N}{N/2+j}, \label{Hsspectrum}
\eea where $0\le j \le N/2$.
Approximating the binomial in Eq. (\ref{Hsspectrum})
by a Gaussian and taking
the limit $N\rightarrow \infty$
while keeping $T/N$ constant we obtain
$\exs{H_S}\approx 3NT/(T+2N)$ and
$T_E\approx 4N$  which
is in agreement with our numerical calculations.
Thus $T_E$ increases linearly with $N$,
the reason being that the number of two-body interaction terms increases
quadratically with the system size.

\subsection{Bose-Hubbard model}

Consider now a lattice model,
the one-dimensional
Bose-Hubbard model, in which the number of
particles can vary on the lattice sites.
We use the language of second-quantization.
Each lattice site corresponds to a bosonic mode with
a destruction operator $a_k$.
The Hamiltonian is \cite{JB98} \be H_B:=-J
\sum_{\langle k,l \rangle} \ad_k a_l + a_k \ad_l + U \sum_k \ad_k
\ad_k a_k a_k, \ee where $J$ is the inter-site tunneling and $U$
is the on-site interaction. Let us consider the case when there is
at most a single particle per lattice site ($U\gg J$) \cite{P04}. Then,
for separable states the energy is bounded from below as \be
\ex {H_B} \ge E_{B,\text{sep}} := -2JN_b
\left(1-\frac{N_b}{N}\right), \label{HB} \ee where $N$ is the number of
lattices sites and $N_b:=\exs{\sum_k \ad_k a_k}$ is the
number of bosonic particles.
For $N=10$ and $N_b=N/2$ (half filling) we obtain
$T_E\approx 0.69J$.

Eq. (\ref{HB}) can be proved as follows. Let us
consider a site in a pure state $\ket{\Psi}=\alpha\ket{0}+\beta\ket{1}$
such that $|\alpha|^2+|\beta|^2=1$.
For this single-site state $|\exs{a}|=|\alpha\beta|$ and $\ex{\ad
a}=|\beta|^2$. Hence $|\exs{a}|^2 = \exs{\ad a}(1-\exs{\ad
a})$. Now using $\sum_{\langle k,l \rangle} \exs{\ad_k}
\exs{a_l} + h.c.\le 2 \sum_k |\exs{a_k}|^2$ one can show
that $E_{B,\text{sep}}$ is an energy bound for product states.
It is a bound also for mixed
separable states of the form Eq.~(\ref{sep}) since
$E_{B,\text{sep}}(N_b)$ is a convex function.

Remarkably, the energy minimum for separable states equals the
minimum for translationally invariant product states. In other
words, it equals the energy minimum obtained from the Gutzwiller
ansatz \cite{KC92} if the expectation value of particle number is
constrained to $N_b$. Note that for our calculations we assumed
that there is at most a single atom per lattice site.

\subsection{Physical realization}

The above methods can be
used for entanglement detection in the following ways:
(i) Energy can be directly measured in some systems
(e.g., optical lattices of cold atoms \cite{JB98}
when used to realize the Bose-Hubbard model).
(ii) The temperature can be measured and used
for entanglement detection.
(iii) The expectation value of the Hamiltonian can be
obtained indirectly if the correlation
terms of the Hamiltonian are measured.
For example, average correlations
$\sum_k\exs{\sigma_a^{(k)}\sigma_a^{(k+1)}}/N$; $a=x,y,z$ can be
measured in a Heisenberg chain
realized with two-state bosonic atoms \cite{GD04}.
From these correlations $\exs{H_H}$ can be computed.

\section{Conclusion}In summary,
we used the Hamiltonian for witnessing entanglement
in spin models. We also considered bosonic lattices.
Our further results concerning this system
will be presented elsewhere \cite{ToBePublished}.
While our method works for non-equilibrium systems,
we have shown that entanglement can efficiently be detected by measuring
energy in a thermal equilibrium.

{\it Note added.---} We presented the idea of using Hamiltonians
as entanglement witnesses on a poster during the Gordon conference in Ventura, 
USA in February 2004. 

{\it Acknowledgment.---}We thank J.I. Cirac, A.C. Doherty, O.
G\"uhne, P. Hyllus, V. Murg and M.M. Wolf for useful discussions.
We
thank J.J. Garc\'{\i}a-Ripoll for
helpful
discussions on
the Bose-Hubbard model, and K. Hammerer for suggesting
Ref.~\cite{CE99}. We
acknowledge the support of the European
Union (Grant No. MEIF-CT-2003-500183), the EU projects RESQ and
QUPRODIS.
and the Kompetenznetzwerk
Quanteninformationsverarbeitung der Bayerischen Staatsregierung.



\begin{thebibliography}{99}
\bibitem{QUANTUMCLASICAL} $\mathcal{O}$ is
a sum of products of spin coordinates. These products have the
form $\prod_k s_{a_k}^{(k)}$ where $a_k \in \{0,x,y,z\}$ and
$s_0^{(k)}=1$. Thus every spin can appear at most once. These
restrictions are necessary if we want to use $\mathcal{O}$ for
getting the minimum for $\exs{O}$. For example,
$\mathcal{O}=(s_x^{(1)})^2$ is not of the required form. Its
minimum is zero, while the corresponding quantum operator is
$O=(\sigma_x^{(1)})^2=\eins$.
\bibitem{SPIN1} Although we consider the spin-1/2 case,
our approach can straightforwardly be generalized for
spin-1 lattices.
\bibitem{ODDEVEN} If the number of sites along one dimension is odd,
then a lattice with periodic boundary
conditions cannot be partitioned into two sublattices such that
neighboring sites correspond to different sublattices. In
this case $E_{\text{sep}}$
given in this paper
is
still a lower bound for the energy of separable states, but
not necessarily the highest possible lower bound.
\end{thebibliography}
\end{document}